\documentclass[useAMS,usenatbib]{mnras}
\usepackage{booktabs}
\usepackage{amsmath}
\usepackage{float}
\usepackage{graphicx}
\usepackage{txfonts}
\usepackage{indentfirst}
\usepackage{color}
\usepackage{ulem}
\usepackage{hyperref}
\usepackage[Symbolsmallscale]{upgreek}
\usepackage{nomencl}
\usepackage{tabularx}
\usepackage[table,xcdraw]{xcolor}
\usepackage{subcaption}
\usepackage{mathbbol}

\newcommand{\eq}[1]{\begin{equation}  #1 \end{equation}}
\newcommand{\eqs}[1]{\begin{equation} \begin{split} #1 \end{split} \end{equation}}

\newcommand{\br}[1]{\left( #1 \right)}
\newcommand{\bc}[1]{\left\{ #1 \right\}}
\newcommand{\bb}[1]{\left[ #1 \right]}

\newcommand{\dd}{{\rm d}}
\newcommand{\expo}[1]{~{\rm e}^{ #1 }}
\newcommand{\vek}[1]{\mbox{\boldmath $#1$}}

\newcommand{\svek}[1]{\mbox{\boldmath \scriptsize $#1$}}

\newcommand{\ic}{{\rm i}}

\def\araa{ARA\&A}
\def\apj{ApJ}
\def\apjl{ApJ}

\def\mnras{MNRAS}
\def\nat{Nature}

\title[Morphology of plasma lenses]{Morphology of solar system scale plasma lenses in the interstellar medium: a test from pulsar scintillation parabolic arcs}

\author[Shi]{Xun Shi$^{1}$\thanks{E-mail: xun@ynu.edu.cn} \\
$^{1}$South-Western Institute for Astronomy Research (SWIFAR), Yunnan University, 650500 Kunming, P. R. China}

\begin{document}
\maketitle
\label{firstpage}
\begin{abstract}
Scintillation spectra of some pulsars have suggested the existence of $\lesssim$ AU scale density structures in the ionized interstellar medium, whose astrophysical correspondence is still a mystery. The detailed study of Brisken et al. suggested two possible morphologies for these structures: a parallel set of filaments or sheets (the `parallel stripes model'), or a filament broken up into denser knots (the `threaded beads model'). 
Here we propose a straightforward test that can distinguish these two morphologies: whether the apex of the main parabolic arc created by the scattered images deviates from the origin of the scintillation spectrum or not. In the `parallel stripes' model, the scattered images move along the stripes as the relative position of the pulsar moves. As a result, the pulsar is always co-linear with the scattered images, and thus the apex of the main parabolic arc stays at the origin of the scintillation spectrum. In the `threaded beads' model, the scattered images remain at almost fixed positions relative to the density structures, and thus the pulsar is not co-linear with the scattered images at most times, leading to an offset between the apex and the origin. Looking for this possible offset in a large sample of pulsar scintillation spectra, or monitoring the evolution of parabolic arcs will help pin down the morphology of these tiny density structures and constrain their astrophysical origin.

\end{abstract}


\section{Introduction}
Pulsar scintillation is the best probe for the tiny, $\lesssim$ AU scale structures of the ionized interstellar medium (IISM). Density inhomogeneities in the IISM cause radio waves originally traveling in different directions to interfere, producing frequency structures in pulsar spectra which vary in space.  
The relatively high transverse velocities of most pulsars effectively scan this spatial variation, leading to pulsar dynamic spectra that vary significantly on timescales of minutes to hours. In the early days of pulsar scintillation studies, pulsar dynamic spectra statistics were used to measure the electron density power spectrum in the interstellar medium on $\sim$AU and smaller scales (see \citealt{rickett90} for a review), and thus played an important role in the discovery of the ``Big Power Law in the Sky'' \citep{armstrong95} -- that the electron density power spectrum follows a single power-law shape across many orders of magnitude in scale \citep[see also][]{chepurnov10, lee19}. 

More recent studies have discovered discrete power distributions on the secondary spectra i.e. square amplitude of Fourier-transformed dynamical spectra. 
First, the power distribution is found to concentrate on a distinct parabolic arc \citep{stinebring01}, and sometimes multiple parabolic arcs \citep{putney06}. These parabolic arcs correspond to the scattered light in contrast to the undiverted light from the pulsar. The thinness of many observed parabolic arcs suggests the existence of discrete scattering screens between the pulsar and the observer, and that the scattering is highly anisotropic \citep{walker04, cordes06}. 
Then, inverted arclets on top of the main parabolic arc have been discovered \citep{hill03}. Weeks-long monitoring of the evolution of the arclets reveals that they move along the main parabolic arc at a fixed speed explainable by the proper motion of the pulsar \citep{hill05}, i.e., the scattered images are long-lived and reside at almost fixed positions in the sky. These imply the existence of $\lesssim$AU scale compact structures in the interstellar plasma that act as plasma lenses \citep{brisken10, pen14, liu16}. 

It is not yet clear how frequent arclets occur, and thus how universal are the corresponding compact plasma lenses in the interstellar medium. The main parabolic arc has been detected in almost every pulsar that has sufficient brightness cf. the GBT Scintarc Survey (Stinebring et al. 2021 in prep, see \citealt{rickett21}). On contrary, arclets have been reported in only a handful of pulsars and have been studied in detail only for a single pulsar B0834+06 \citep{hill05, brisken10}. However, in the high dynamic range study of pulsar B0834+06 \citep{brisken10}, the main parabolic itself is resolved into numerous discrete inverted arclets, implying the possibility that these tiny plasma lenses are the major contributor of the main parabolic arcs, and thus the possible ubiquity of these lenses. The same tiny lenses may also be responsible for the extreme scattering events (ESEs) of radio quasars and pulsars \citep[e.g.][]{fiedler87, fiedler94,bannister16, kerr18}.

What are these tiny, compact, and likely ubiquitous plasma structures remains a mystery.
Due to the lack of other observational probes on such small physical scales, we rely completely on scintillation studies to infer the properties of these plasma lenses. One clue scintillation studies can offer is the morphology of the plasma lens, as scattering material with different orientation and relative velocity with respect to the source would yield different secondary spectra (see e.g. appendix of \citealt{xu18} for a demonstration).
\citet{brisken10} proposed two models for the possible morphology of the scattering material in front of pulsar B0834+06 during their observation: (a) a roughly parallel set of filaments or sheets, with each filament or sheet being the underlying lens for one arclet; and (b) a filament with denser knots, with each knot being the underlying lens for one arclet. These different morphologies are indicative of their astrophysical origin. The morphology (a) can be produced by a corrugated reconnection sheet \citep{pen12, pen14b, liu16, simard18}, a parallel set of shock fronts or ionization fronts, or a parallel set of filaments tracing the magnetic fields;
On the other hand, the morphology (b) can be produced by a filament ionized by hot stars, a projected edge of a corrugated supernova shell, an ionized skin of a molecular clump \citep{walker17}, or structures akin to a flux rope in the solar wind \citep{brisken10}.

We demonstrate that one can distinguish the two morphologies by examining whether the apex of the main parabolic arc aligns with the origin on pulsar secondary spectra. Namely, the apex of the main parabolic arc stays at the origin of the secondary spectrum in model (a), but can move away from the origin in model (b) as the relative position of the pulsar changes with time. We present our analytical models for (a) and (b) in Sect.\;\ref{sec:geo}, show how the corresponding pulsar secondary spectra evolve differently in these scenarios in Sect.\;\ref{sec:sec}, discuss the implications from current data in Sect.\;\ref{sec:discussion} and conclude in Sect.\;\ref{sec:con}.

\section{Modeling}
\label{sec:geo}

\begin{figure}
    \centering
        \includegraphics[width=0.4\textwidth]{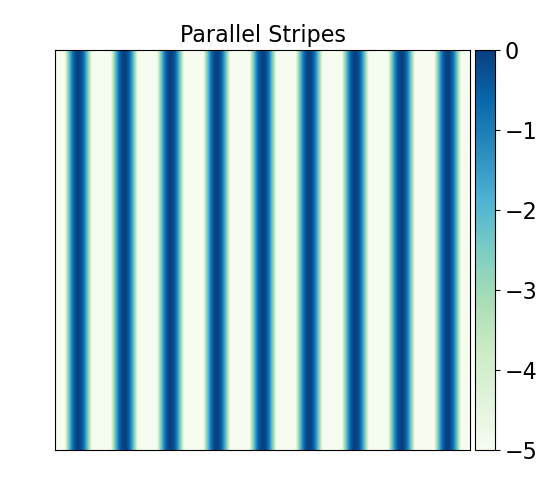}   \\ 
        \includegraphics[width=0.4\textwidth]{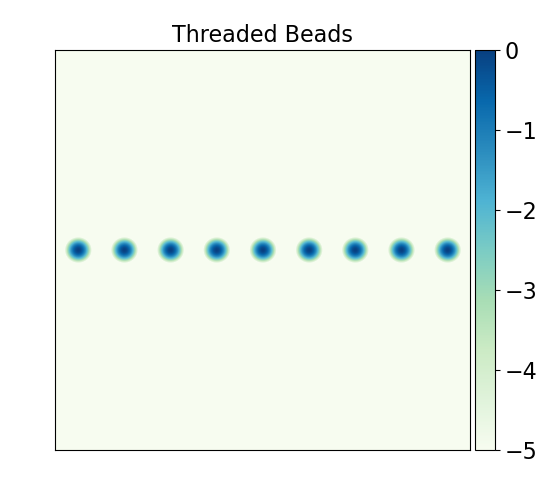}     
        \caption{\textbf{Lensing potentials of the parallel stripes model (upper panel) and the threaded beads model (lower panel).} Both the sheet and the filament of high plasma densities are corrugated in the horizontal direction, composed of many individual high-density regions (sub-lenses). Whereas each sub-lens is extended in the vertical direction (i.e. like a stripe) in the parallel stripes model, it is confined in both directions (i.e. like a knot) in the threaded beads model. The lensing potential of a single sub-lens is  described by Eq.\;\ref{eq:phi}. The colorbar indicates the overdensity of the plasma in logscale and normalized to the maximum value.}
        \label{fig:phimap}
\end{figure}

\begin{figure*}
    \centering
        \includegraphics[width=0.9\textwidth]{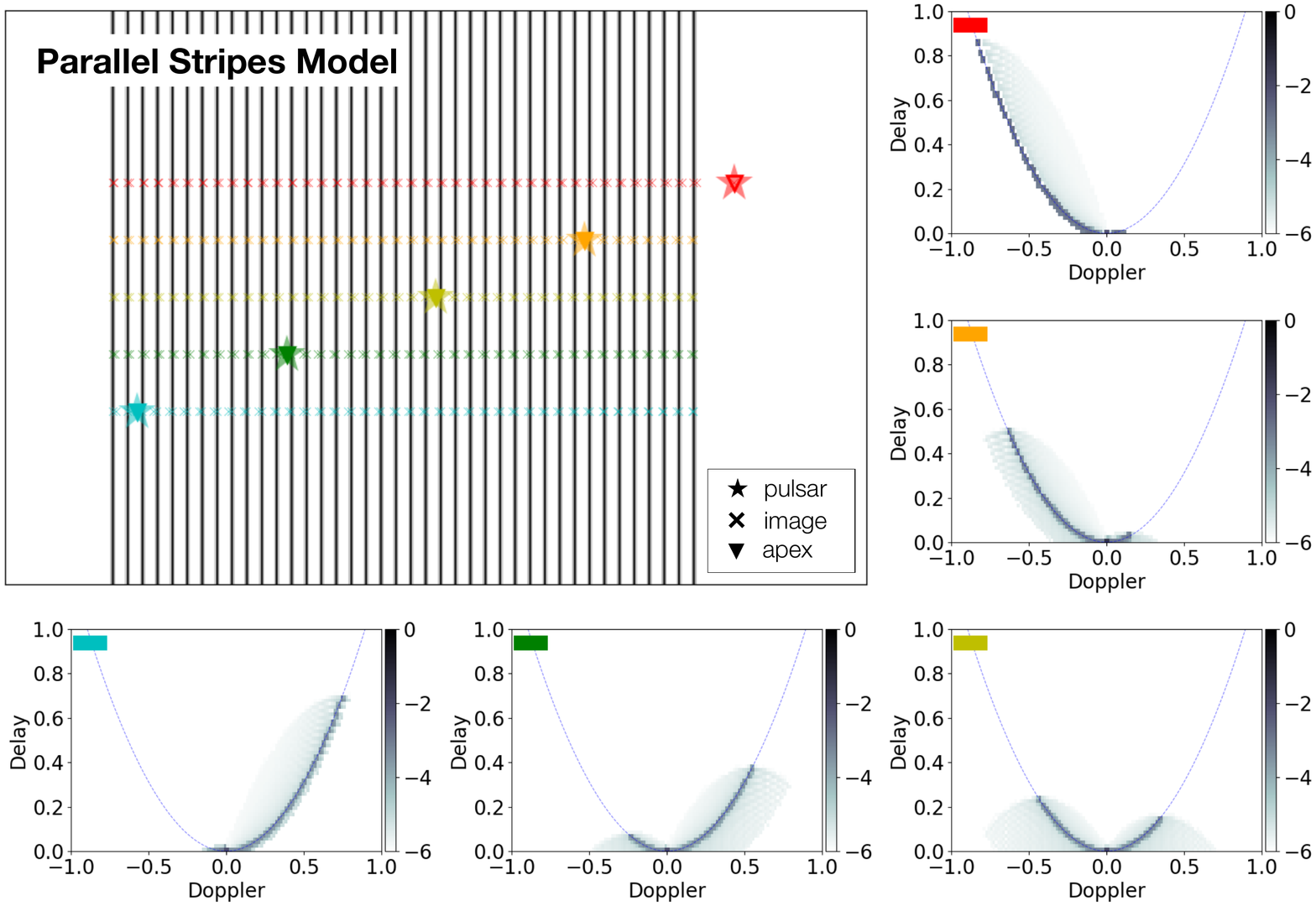}        
        \caption{\textbf{Configuration and secondary spectra of pulsar traveling behind a scattering screen described by the parallel stripes model (upper panel of Fig.\;\ref{fig:phimap}).} As a pulsar travels behind a foreground scattering screen in the form of a corrugated sheet, it creates multiple images marked as  crosses in the upper left panel. The images include a main image at approximately the angular position of the pulsar, and scattered images close to the plasma density peaks of the sub-lenses shown as the gray vertical stripes. We have adopted forty sub-lenses covering a finite spatial region. Secondary spectra of pulsar scintillation (small panels) are computed for five pulsar angular positions, with the color marked on a secondary spectrum matching that of the star in the upper left panel indicating the pulsar position. 
        The main parabolic arc on a secondary spectrum is produced from the interference of the scattered images with the main image. Its apex corresponds to an image position indicated with a triangle of the corresponding color in the upper left panel. The doppler and delay axes are normalized to arbitrary units that are the same for all secondary spectra. The blue dashed lines on different secondary spectra show the same parabola to guide the eye.  {For the parallel stripes model, the angular position of the pulsar is always collinear with the line of scattered images. Therefore, the apex position of the main parabolic arc stays at the origin of the secondary spectrum.}}
        \label{fig:sheet} 
\end{figure*}

\begin{figure*}
    \centering
        \includegraphics[width=0.9\textwidth]{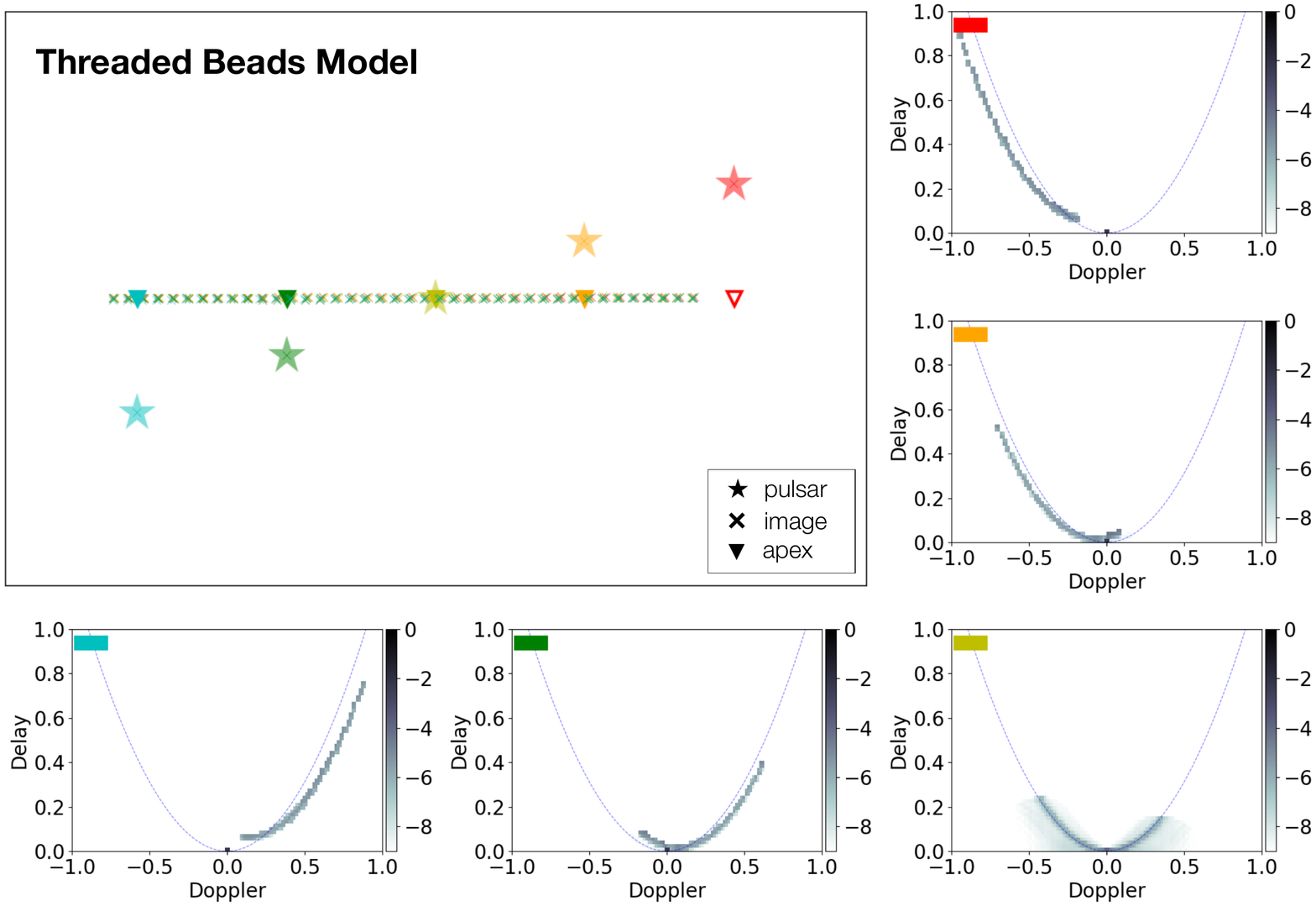}        
        \caption{\textbf{Same as Fig.\;\ref{fig:sheet} but for the threaded beads model (lower panel of Fig.\;\ref{fig:phimap}).} The sub-lenses are not clearly visible in the upper left panel of this plot due to their small sizes, but their positions are centered roughly at those of the scattered images. {For the threaded beads model, the angular position of the pulsar can have a large deviation from the line of scattered images. As a consequence, the apex position of the main parabolic arc can have a large deviation from the origin of the secondary spectrum. }} 
        \label{fig:filament} 
\end{figure*}

\subsection{Plasma Lens Models}
\label{sec:lens_model}
In the eikonal limit, the physical mechanism of radio wave deflection in the ionized interstellar medium is akin to that of gravitational lensing \citep{clegg98, pen12,er18,wagner20,shi21}. Accordingly, we regard the scattering screen as a plasma lens, and the tiny compact density structures responsible for the arclets as sub-lenses.

Our central goal is to distinguish the two morphological models of the projected potential of the plasma lens  that were suggested by \citet{brisken10}: (a) a corrugated sheet composed of sub-lenses in the shape of stripes, and (b) a corrugated filament composed of sub-lenses in the shape of knots (Fig.\;\ref{fig:phimap}). Following the framework described in \citet{shi21}, we quantify the light-bending ability of a plasma lens with its lensing potential, and model the lensing potential $\psi$ of a single sub-lens with a simple Gaussian form
\eq{
\label{eq:phi}
 \psi(\theta) = A  \expo{- \theta^2 / \theta_0^2}   \,,
}
where $\theta_0$ is the characteristic size of the sub-lens and $A$ the lens amplitude. The variable $\theta = \theta_x$ for the parallel stripes model, and $\theta = \sqrt{\theta_x^2 + \theta_y^2}$ for the threaded beads model, $\theta_x$ and $\theta_y$ being the two cartesian angular coordinates on the plane of the sky. 

Physically, the lens amplitude $A$ is set by a combination of the column density of the free electrons in the scattering screen, the frequency of the radio wave, and the distances (see e.g. Eq.\;7 of \citealt{shi21}). It determines the amplitude of the dispersive phase shift caused by the plasma lens, and the ratio of $A/\theta_0$ constrains the size of the deflection angle. Here, we choose a lens amplitude $A \approx 10^3 \theta_0^2$ so that it lies in the regime $\theta_0^2 \ll A \ll \rm{typical\ deflection\ angle\ squared}$. We think this regime applies to the arclets. In this regime, the lens can create local images even when its angular position is far away from that of the source, and thus can explain the existence of arclets at large doppler values. At the same time, the
dispersive phase shift introduced by this lens amplitude is typically less than a few percent of the relative geometric delay of an image scattered by a sub-lens with respect to the undiverted main image, consistent with the dominance of the geometric delay as suggested by observations. 

The characteristic size of the sub-lens $\theta_0$ is chosen to be one order of magnitude smaller than the separation between sub-lenses to ensure the discreteness of the sub-lenses. The size of $\theta_0$ itself is a free parameter in our models with no effect on the results. Past observational estimates constrain it to be on AU to sub-AU scales. We consider forty equally-spaced sub-lenses covering a finite spatial region as a demonstration in Figs.\;\ref{fig:sheet} and \ref{fig:filament}.

\citet{brisken10} named the models (a) the `orthogonal' model and (b) the `parallel' model. Here, we rename them as (a) the `parallel stripes model' and (b) the `threaded beads model' to describe their key difference -- their own morphology instead of their alignment with the moving direction of the pulsar. To show that the latter does not play a major role, we let the pulsar travel at a 30-degree angle with respect to the x-direction (see upper left panel of Figs.\;\ref{fig:sheet} and \ref{fig:filament}), i.e., the direction of the relative velocity is neither in parallel with nor orthogonal to the symmetry directions of the plasma density distribution. 

The magnifications $\mu$ of the sub-images are set by $\mu = 1 / \rm{det}\br{\mathbb{1} - \partial_{\svek{\theta}} \partial_{\svek{\theta}} \psi}$, where $\mathbb{1}$ is the identity matrix  (see Eqs.\;11 and 12 of \citealt{shi21} for details). For localized sub-lenses in the parameter regime described above, $|\mu| \approx 1 / |\rm{det}\br{\partial_{\svek{\theta}} \partial_{\svek{\theta}} \psi}|$. For the parallel stripes model where $\psi$ is constrained only in one of the two cartesian dimensions, the magnifications of the arclets are on the order of $A / \theta_0^2$; and for the threaded beads model, they are of order $(A / \theta_0^2)^2$ -- for both models we have $|\mu| \ll 1$. These are consistent with the relative brightness of the observed arclets at large doppler: in \citet{brisken10}, the main parabolic arc (i.e. the apex of the arclets) at $ \gtrsim 30$ mHz is three orders of magnitude dimmer than its peak brightness, and similar brightnesses were observed in \citet{hill05}. However, our models with the chosen parameter range cannot explain the gradual brightening of the arclets as they approach the origin on the secondary spectrum. This `gradual brightening' is hinted in the arclet brightness distribution in \citet{brisken10} and later confirmed by the multi-epoch observations of PSR B0834+06 (Simard et. al. in prep) and PSR B0355+54 (Yao et. al. in prep). In our models, the brightening should occur only abruptly near the location of the inner caustics. Note that, although some density fluctuation distributions on the scattering screen (e.g. an anisotropic Kolmogorov distribution) can qualitatively reproduce the brightness distribution on the main parabolic arc \citep[e.g.][]{walker04, cordes06}, their explanation for the `brightening towards the origin' is based on statistically more radio waves deflected at smaller angles, and thus they do not explain the brightening of a single arclet either. This `gradual brightening' remains a puzzle that may provide a key clue to the nature of the scattering that leads to the arclets. For now, in the following of this paper, we shall focus on the morphology of the scattering alone.

\subsection{From Lens to Secondary Spectrum}
We approximate the wavefield after passing through the scattering screen as contributions from a finite number of discrete image points where the phases are stationary \citep[e.g.][]{walker04}.
This stationary phase approximation holds well for regions away from the caustics when the separations in consideration are much greater than the Fresnel scale \citep{jow20, shi21}. Under this approximation, we can apply the geometrical lensing framework, and derive how many images the plasma lens produces, as well as the angular position $\vek{\theta}_i$ and magnification $\mu_i$ of each image $i$ from the plasma lens model given the angular position of the source $\vek{\beta}$.

We then adopt the electric field representation as introduced by \citet{walker05} for the pulsar intensity spectra, i.e. the received electric field $E$ of the pulsar is decomposed into contributions from all images. Up to an arbitrary normalization,
\eq{
E(\nu, t) = \sum_i u_i \exp \bb{2 \uppi \ic \br{\omega_i \br{t - t_0} - \tau_i \br{\nu - \nu_0}}} \,,
}
where time $t_0$ and frequency $\nu_0$ are the central values of an observation. The electric field from an image $i$ is characterized by its doppler 
\eq{\omega_i = \frac{(\vek{\theta}_i - \vek{\beta}) \cdot \vek{V}_{\rm eff}}{\lambda} \,,
} 
delay 
\eq{
    \tau_i = \frac{|\vek{\theta}_i - \vek{\beta}|^2 D_{\rm eff}}{2 \it{c}} \,,
} together with its amplitude $u_i \propto \sqrt{|\mu_i|}$ taken at $t=t_0$ and $\nu=\nu_0$. For a scattering screen located at a fractional distance $f_{\rm d}$ from the pulsar to the observer, $D_{\rm eff} \equiv D_{\rm p} (1 - f_{\rm d}) / f_{\rm d}$ with $D_{\rm p}$ being the pulsar distance, and
$\vek{V}_{\rm eff} \equiv {\vek{V}}_{\rm sr} / f_{\rm d}$ with ${\vek{V}}_{\rm sr}$ being the relative velocity of the scattering screen with respect to the pulsar-observer line of sight  \citep[see e.g.][]{brisken10}. When the relative velocity is dominated by the proper motion velocity $V_{\rm p}$ of the pulsar, ${\vek{V}}_{\rm sr} = -\vek{V}_{\rm p} (1 - f_{\rm d})$. 

A dynamic spectrum records the variation of the pulsar intensity $I = E E^*$ with time and frequency,
\eq{
I(\nu, t) =  \sum_{i, j} u_i u_j \exp \bc{2 \uppi \ic \bb{ \br{\omega_i - \omega_j} \br{t - t_0} - \br{\tau_i - \tau_j}\br{\nu - \nu_0}}}  \,.
} 
The corresponding secondary spectrum $S = \tilde{I}\tilde{I}^*$ is
\eq{
\label{eq:sec}
S(\omega, \tau) = \sum_{i, j} u_i^2 u_j^2 \delta_D \br{\omega - \omega_i + \omega_j} \delta_D \br{\tau + \tau_i - \tau_j} \,,
}
with $\delta_D$ being the Dirac delta function. According to Eq.\;\ref{eq:sec}, we compute the secondary spectrum by summing over all image pairs, with each image pair ($i,j$) contributing a power proportional to $u_i^2 u_j^2 \propto |\mu_i \mu_j|$ at a position described by its doppler $\omega = \omega_i - \omega_j$ and delay $\tau = -\tau_i + \tau_j$. Finally, we pixelize this secondary spectrum using 128 pixels along each axis. 

Examples of the resulting secondary spectra are given in the small panels in Figs.\;\ref{fig:sheet} and \ref{fig:filament}) for the parallel stripes model and the threaded beads model, respectively. Secondary spectra for both models show clear main parabolic arcs from the interference between the main image and the scattered images. Extended arclets with their apexes lying on the main parabolic arc arise from interference among the scattered images. They are visible in all panels in Fig.\;\ref{fig:sheet} for the parallel stripes model. For the threaded beads model, only one panel on the bottom right of Fig.\;\ref{fig:filament} shows extended arclets above 80 dB. This distinction is a result of the fainter images associated with the sub-lenses in the threaded beads model, a natural outcome of the lensing potential shapes.

The point-like, undiverted main image at the source angular position $\vek{\beta}$ and the well-separated sub-lenses we adopt in this paper make it easy to demonstrate the relative positions of the pulsar and the images which is the main purpose of this paper, but it is unrealistic in the sense that the scintillation is weak, in conflict with the observations. In reality, the scattering screen can create a whole spectrum of images of various brightnesses and separations. There is no clear distinction between the main image and the scattered images, and the most notable discrete arclets are just the correspondents of distinct bright images well separated from the source.  

\section{Evolution of Secondary Spectrum}
\label{sec:sec}

\begin{figure}
    \centering
        \includegraphics[width=0.34\textwidth]{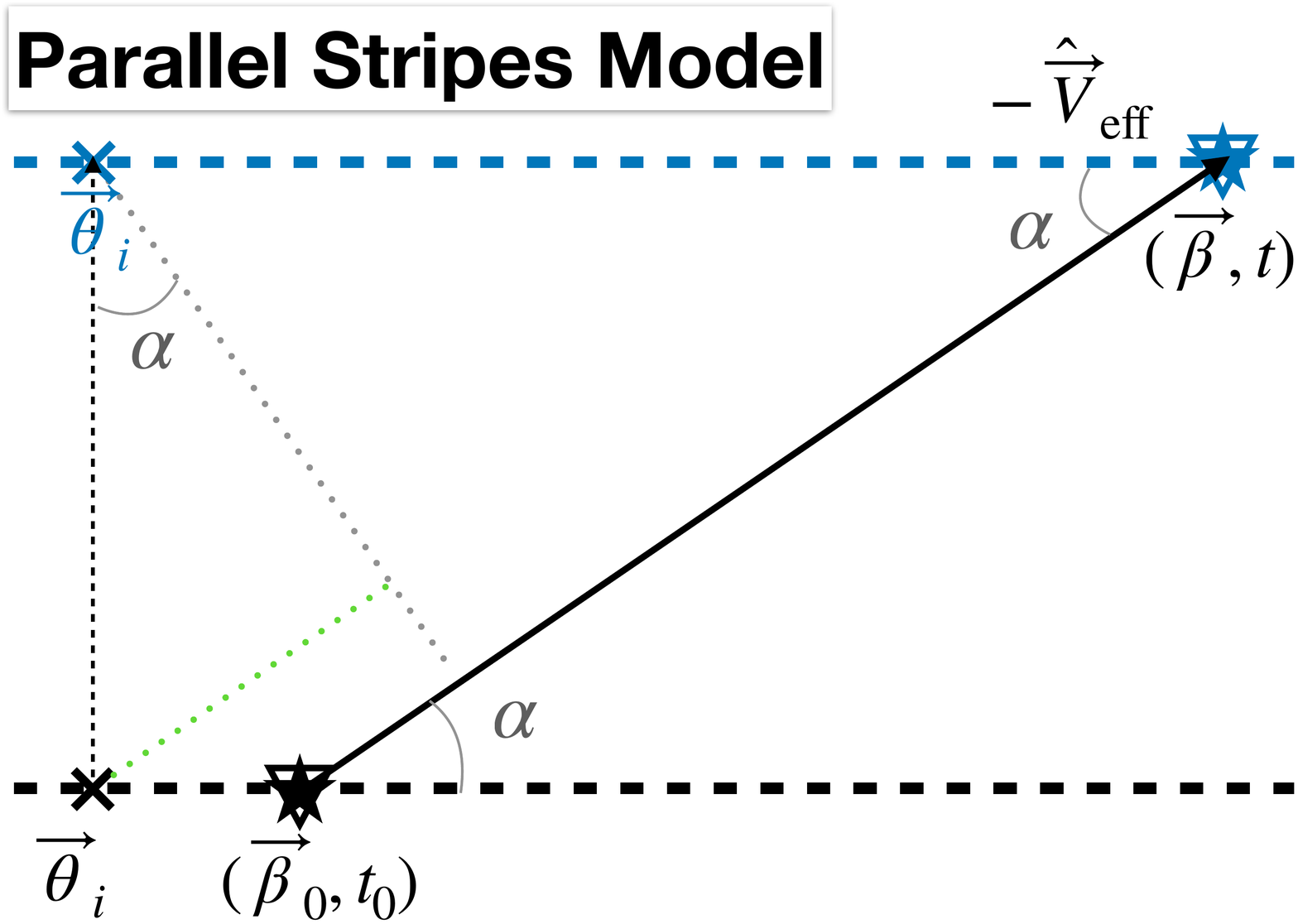}   \\  
        \includegraphics[width=0.34\textwidth]{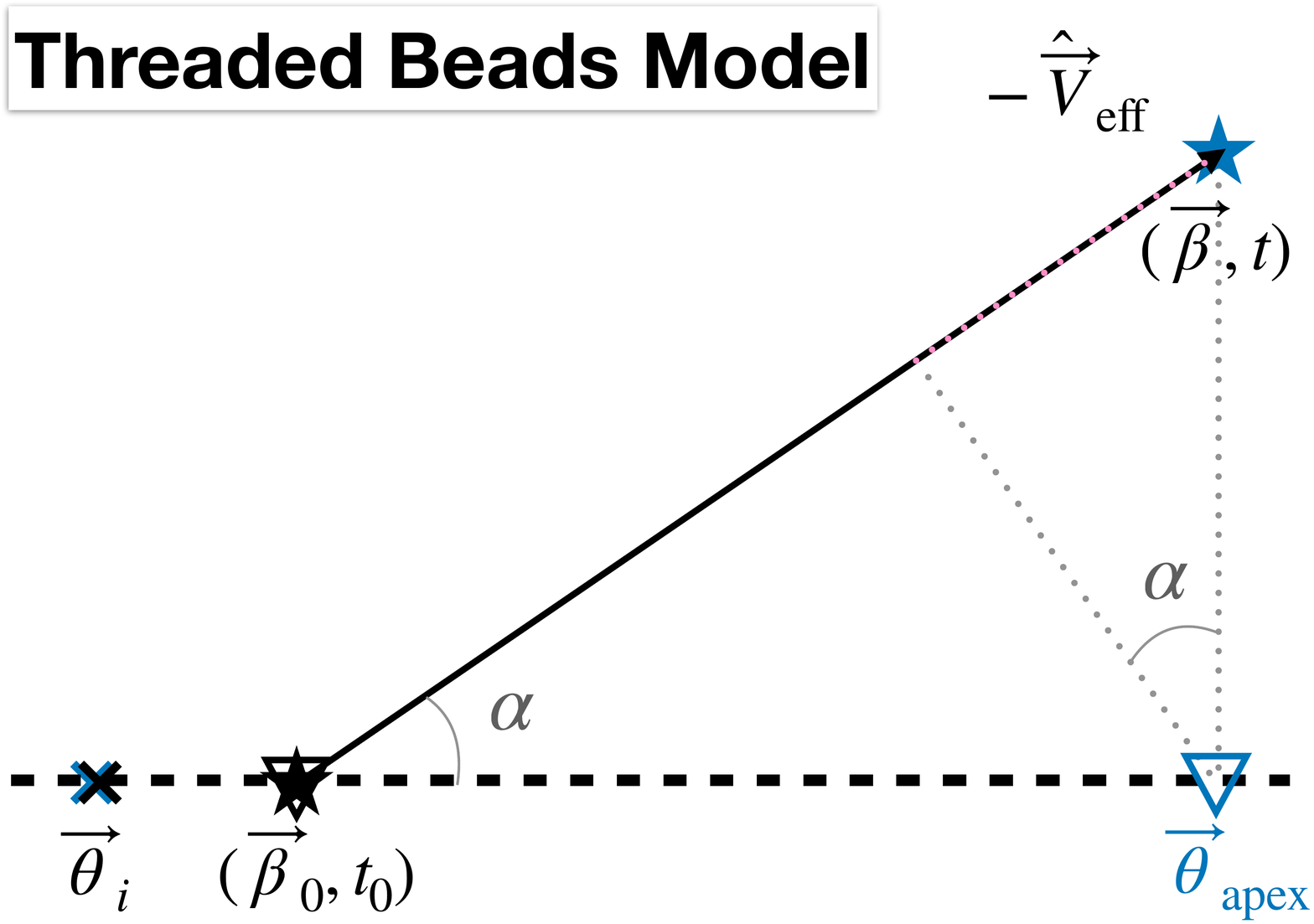}   
        \caption{\textbf{Geometries of relative motions of the pulsar (`$\bigstar$'), the apex of the main parabolic arc (`$\nabla$'), and one scattered image (`$\times$')} for the parallel stripes model (upper panel) and the threaded beads model (lower panel). The line of scattered images is marked as the dashed line. Relative positions at time $t_0$ when the pulsar co-aligns with the line of scattered images are indicated with the black symbols, whereas the blue symbols stand for the positions at another time $t$.} 
        \label{fig:geo} 
\end{figure}

\begin{table*}
    \caption{\textbf{Motion of apex of the main parabolic arc and the scattered images on the secondary spectrum (Doppler $\omega$ - delay $\tau$) caused by the relative motion between the pulsar and the plasma lens.} 
    For the threaded beads model, $t_0$ is the time of the special moment when the pulsar is co-linear with the line of scattered images. Note that we have neglected the slight shifts of the image locations with respect to the sub-lens centers. 
   }
    \begin{tabular}{@{}c|c|c|c|c@{}}
        \toprule
                                & apex doppler \textbf{$\omega_{\rm apex}$}                                      & apex delay \textbf{$\tau_{\rm apex}$}                                             & image dopper speed \textbf{$\rm d \omega_i / \rm d t$}                         & image delay speed \textbf{$\rm d \tau_i / \rm d t$}   \\ \midrule
        Parallel stripes model    & 0                                                                 & 0                                                                      & $\frac{{V}_{\rm eff}^2  \cos^2\alpha}{\lambda D_{\rm eff}}$ & $\frac{\lambda}{ \it{c}} \omega_i$ \\ \midrule
        Threaded beads model & $\frac{V_{\rm eff}^2 \sin^2\alpha}{\lambda D_{\rm eff}}(t - t_0)$ & $\frac{{V}_{\rm eff}^2 \sin^2\alpha}{2 \it{c} D_{\rm eff}}(t - t_0)^2$ & $\frac{{V}_{\rm eff}^2}{\lambda D_{\rm eff}}$               & $\frac{\lambda}{ \it{c}} \omega_i$ \\ \bottomrule
    \end{tabular}
    \label{tab:1}
\end{table*}

A secondary spectrum carries direct information about the images created after light deflection at the scattering screen. Generically, the images consist of a main image close to the angular position of the pulsar that has a brightness close to the original pulsar brightness i.e. $\mu \approx 1$, and scattered images that are much fainter $\mu \ll 1$. In such a case, the self-interference of the main image leads to the highest power locating at the origin of the secondary spectrum, the interference between the main image and the scattered images leads to the fainter main parabolic arc, and that among the scattered images leads to the even fainter inverted arclets. Here, we do not consider the rare case when a sub-lens transits the pulsar, or when scattered images are produced or annihilated at a caustic (see e.g. \citealt{shi21} for such a case).  

The relative position of the main parabolic arc apex to that of the origin on a secondary spectrum reflects the relative angular positions of the scattered images to that of the pulsar in the sky. In the parallel stripe model, a stripe deflects radio waves in a direction perpendicular to the stripe but not in a direction parallel to the stripe; so that the image from a particular stripe appears to move along the stripe as the pulsar moves (Fig.\;\ref{fig:sheet}). In contrast, 
the threaded beads model produces scattered images that are local to the filament location and stay at fixed positions in the sky as the pulsar moves (Fig.\;\ref{fig:filament}) \footnote{We neglect the slight shifts of the image positions with respect to the lens as the relative position of the pulsar changes, which only create perturbative effects to the geometries we demonstrate in this paper.}. As a result, the pulsar is always co-linear with the scattered images for the parallel stripes model, and is co-linear only at one specific time for the threaded beads model. 

The apex of the main parabolic arc at an angular position $\vek{\theta}_{\rm apex}$ has the minimum delay among all points on the main parabolic arc, and is created by the point on the line of the scattered images that is the closest to the pulsar. When the pulsar and the scattered images are not co-linear, there is a positive minimum delay of the scattered images with respect to the pulsar, i.e. the delay at the apex of the main parabolic arc $\tau_{\rm apex} > 0$, the apex moves away from the origin. Thus, whereas the apex of the main parabolic arc always stays at origin on the secondary spectrum if the scattering screen is described by the parallel stripes model, it typically deviates from the origin if the scattering screen is described by the threaded beads model. This offers a powerful test of the scattering screen morphology.

To figure out quantitatively how the apex and the inverted arclets (i.e. the scattered images) move on the secondary spectrum, we work in a coordinate in which the plasma lens location is fixed, and the pulsar location $\vek{\beta}$ moves with time in the direction of $ - \vek{V}_{\rm eff}$ relative to the lens: $\vek{\beta} - \vek{\beta}_0  = - \vek{V}_{\rm eff} (t - t_0) /  D_{\rm eff}$.
Thus, 
\eq{
    \frac{\dd {\vek{\beta}}}{\dd t} =  - \frac{\vek{V}_{\rm eff}}{D_{\rm eff}} \,.
}
The time $t_0$ is chosen as a time when the pulsar is co-linear with the scattered images for the threaded beads model, and is arbitrarily chosen for the parallel stripes model. Fig.\;\ref{fig:geo} shows the geometries of the relative motions, with the line of scattered images shown as the dashed line, and the angle between this line and the direction of the relative motion of pulsar denoted as $\alpha$. 

\subsection{Parallel stripes model}
For the parallel stripes model, the main image at the angular position of the pulsar $\vek{\beta}$ always stays co-linear with the scattered images. Consequently, $\vek{\theta}_{\rm apex} = \vek{\beta}$, and thus $\omega_{\rm apex} = 0$, $\tau_{\rm apex} = 0$. The angular position shift of a scattered image $\Delta \vek{\theta}_i$ projects onto the direction of $\vek{V}_{\rm eff}$ has a magnitude $\Delta \vek{\theta}_i \cdot \hat{\vek{V}}_{\rm eff} = \Delta{{\beta}} \sin^2 \alpha$ (cf. green dotted line in Fig.\;\ref{fig:geo}). 
The relative position of the scattered images with respect to the pulsar $\br{\vek{\theta}_i - \vek{\beta}}$ stays aligned with the line of scattered images and is perpendicular to $\dd \vek{\theta}_i / \dd t$.
Thus, the speed of the arclet motion in Doppler and delay are
\eq{
\frac{\dd \omega_i}{\dd t} \equiv \frac{1}{\lambda} 
\frac{\dd (\vek{\theta}_i - \vek{\beta})}{\dd t} \cdot \vek{V}_{\rm eff} =  -\frac{\cos^2\alpha }{\lambda} \frac{\dd {\beta}}{\dd t}  {V}_{\rm eff} =  \frac{{V}_{\rm eff}^2  \cos^2\alpha}{\lambda D_{\rm eff}}  \,,
}
and
\eq{
\frac{\dd \tau_i}{\dd t}  \equiv \frac{D_{\rm eff}}{2\it{c}} \br{\vek{\theta}_i - \vek{\beta}} \cdot \frac{\dd (\vek{\theta}_i - \vek{\beta})}{\dd t} =  - \frac{D_{\rm eff}}{\it{c}} \br{\vek{\theta}_i - \vek{\beta}} \cdot \frac{\dd \vek{\beta}}{\dd t}  = \frac{\lambda}{ \it{c}} \omega_i \,.
} 
It stays on the main parabolic arc whose apex is at the origin:
\eq{
\frac{\dd \tau_i}{\dd \omega_i^2} = \frac{\dd \tau_i/\dd t}{2 \omega_i \dd \omega_i / \dd t} = \frac{\lambda^2 D_{\rm eff}}{ 2\it{c} V_{\rm eff}^2 \cos^2\alpha} = \eta \,,
 }
with $\eta \equiv \tau_i / \omega_i^2$ being the curvature of the main parabolic arc.

\subsection{Threaded beads model}
For the threaded beads model,  
the apex of the main parabolic arc corresponds to an angular position $\vek{\theta}_{\rm apex}$ that is the projection of that of the pulsar location onto the line of scattered images (see the blue star and triangle in the lower panel of Fig.\;\ref{fig:geo}). Its Doppler and delay move as the pulsar position $\vek{\beta}$ changes,
\eqs{
    \label{eq:omega_apex_filament}
    \omega_{\rm apex} &\equiv \frac{(\vek{\theta}_{\rm apex} -  \vek{\beta}) \cdot \vek{V}_{\rm eff}}{\lambda} 
     = \frac{V_{\rm eff}^2 \sin^2\alpha}{\lambda D_{\rm eff}}(t - t_0) \,,
}

\eqs{
    \label{eq:tau_apex_filament}
    \tau_{\rm apex} &\equiv \frac{|\vek{\theta}_{\rm apex} - \vek{\beta}|^2 D_{\rm eff}}{2\it{c}} = \frac{|\vek{\beta} - \vek{\beta}_0|^2 \sin^2\alpha  D_{\rm eff}}{2\it{c}}\\
    & = \frac{{V}_{\rm eff}^2 \sin^2\alpha} {2 \it{c} D_{\rm eff}} (t - t_0)^2 \\
    & =  \eta \frac{\cos^2\alpha}{\sin^2\alpha} \omega_{\rm apex}^2 \,.
}
In general, when the effective velocity is not co-linear with the line of sub-images, the apex offsets from the origin: $\omega_{\rm apex} \neq 0$ and $\tau_{\rm apex} > 0$. Similar results have been obtained in the appendix of \citet{xu18}. The trajectory of the apex also forms a parabola centering at the origin. The curvature of this parabola $\eta'$ is different from that of the main parabolic arc $\eta$. The two are related as $\eta' =  \eta \cos^2\alpha / \sin^2\alpha$.  

The positions of the scattered images stay fixed, $\Delta \vek{\theta}_i = 0$ for the threaded beads model, and the arclets move purely as a consequence of the shift of the pulsar position,
\eq{
\frac{\dd \omega_i}{\dd t} = - \frac{1}{{\lambda}}\frac{\dd \vek{\beta}}{\dd t} \cdot \vek{V}_{\rm eff} =  \frac{{V}_{\rm eff}^2}{\lambda D_{\rm eff}}  \,,
}
\eq{
\frac{\dd \tau_i}{\dd t}  = - \frac{D_{\rm eff}}{c} \br{\vek{\theta}_i - \vek{\beta}} \cdot \frac{\dd \vek{\beta}}{\dd t}  =  \frac{ \lambda}{ \it{c}} \omega_i \,.
}

As a summary (see Table.\;\ref{tab:1}), whether the apex of the main parabolic arc offsets from the origin on a secondary spectrum or not can serve as a powerful test for the underlying morphology of the scattering screen. 
Note that there is no physical limit to the size of the offset for the threaded beads model. The only observational constraint is that the parabolic arc should still be bright enough to be observed given the offset. Large offsets are easier to occur when the angle $\alpha$ is large i.e. the direction of the relative velocity is nearly perpendicular to the line of the scattered images (Eqs.\;\ref{eq:omega_apex_filament} and \ref{eq:tau_apex_filament}). In such a case, the curvature of the main parabolic arc $\eta$ is also large, and observationally we would expect a nearly folded parabolic arc with its apex away from the origin.

\section{Discussion}
\label{sec:discussion}

In a recent study \citep{yao21}, a clearly noticeable deviation of the apex of the main parabolic arc from the origin of the secondary spectrum is reported for one pulsar. Moreover, the apex position is found to vary greatly from epoch to epoch, across intervals as short as one day. These are strongly suggestive of a scattering screen described by the threaded beads model. However, this pulsar PSR J0538+2817 likely stands for a special case: it lies within a supernova remnant (SNR), and the study finds that the distance from the pulsar to the near side of the SNR shell is consistent with that of the scattering screen. Thus, in the case of PSR J0538+2817, it is likely that some filamentary structure on the SNR shell has created the parabolic arc on its secondary spectrum. 

In general, a significant offset of the apex from the origin seems to be observationally rare. In the past observations of the few tens of pulsars with parabolic arc detection, we are not aware of any other report of such an offset. Although slight offset may not be detectable given the limited resolution of the secondary spectrum, the possible large offset predicted by the threaded beads model is never observed. This hints at the possibility that a generic scattering screen in the IISM is described by the parallel stripes model. More observations of parabolic arcs, especially monitoring observations of the evolution of a parabolic arc are needed to pin down the morphology of the scattering screen. 

\section{Conclusion}
\label{sec:con}
What are the $\lesssim$ AU scale IISM structures that give rise to the inverted arclets on some pulsar secondary spectrum is an intriguing mystery. Here, as one step towards its solution, we consider the two possible morphological models suggested by \citet{brisken10}. 

We demonstrate that the secondary spectra evolve differently in these two models, i.e., the apex of the main parabolic arc stays at the origin of the secondary spectrum in the parallel stripes model, but can move away from the origin in the threaded beads model. This is because, in the parallel stripes model, the scattered images move with the pulsar which keeps them co-linear with each other; but in the threaded beads model, the locations of the scattered images are almost fixed, and so that the pulsar is not co-linear with the scattered images at most times. We further compute the motion of the apex and the arclets in both models. In particular, in the threaded beads model, the motion of the apex forms a parabolic arc on the secondary spectrum with a curvature different from but related to that of the main parabolic arc.

This distinction offers a direct test of the plasma lens morphology. 
It also calls for more observations of pulsar scintillation arcs looking for the possible offset of the main parabolic apex from the origin on a secondary spectrum. Monitoring observations of the evolution of pulsar parabolic arcs will be especially useful for determining the morphology of these likely ubiquitous $\lesssim$ AU scale IISM structures. 

\section*{Data Availability Statements}
\noindent No new data were generated or analysed in support of this research.

\section*{Acknowledgements}
\noindent XS thanks the referee for his very helpful report, and Yun Lin for discussions about the names of the models.

\bibliographystyle{mnras}
\bibliography{bibliography}

\end{document}